\documentclass[cits]{JINST}
\usepackage{amsmath}

\title{Neganov-Luke amplified cryogenic light detectors for the background discrimination in neutrinoless double beta decay search with TeO$_{2}$ bolometers}

\author{M.~Willers$^a$\thanks{Corresponding author: \texttt{mwillers@ph.tum.de}}~, F.~v.~Feilitzsch$^a$, A.~G\"utlein$^a$\thanks{Present address: Institut f\"ur Hochenergiephysik der \"Osterreichischen Akademie der Wissenschaften, A-1050 Wien, Austria}~, A.~M\"unster$^a$, J.-C.~Lanfranchi$^a$, L.~Oberauer$^a$, W.~Potzel$^a$, S.~Roth$^a$, S.~Sch\"onert$^a$, M.~v.~Sivers$^a$\thanks{Present address: LHEP, Albert Einstein Center, University of Bern, 3012 Bern, Switzerland}~, S.~Wawoczny$^a$, A.~Z\"oller$^a$ and A.~Giuliani$^b$\\
\llap{$^a$}Physik-Department E15 \& Excellence Cluster Universe, Technische Universit\"at M\"unchen, D-85748 Garching, Germany\\
\llap{$^b$}Centre de Sciences  Nucl\'{e}aires et de Sciences de la Mati\`ere, 91405 Orsay Campus, France}

\abstract{We demonstrate that Neganov-Luke amplified cryogenic light detectors with Transition Edge Sensor read-out can be applied for the background suppression in cryogenic experiments searching for the neutrinoless double beta decay of $^{130}\text{Te}$ with $\text{TeO}_{2}$ based bolometers. Electron and gamma induced events can be discriminated from $\alpha$ events by detecting the Cherenkov light produced by the $\beta$ particles emitted in the decay. We use the Cherenkov light produced by events in the full energy peak of $^{208}\text{Tl}$ and by events from a $^{147}\text{Sm}$ source to show that at the Q-value of the neutrinoless double beta decay of $^{130}\text{Te}$ ($Q_{\beta \beta} = 2.53 \,\text{MeV}$), a separation of $e^{-}/\gamma$ events from $\alpha$ events can be achieved on an event-by-event basis with practically no reduction in signal acceptance.}


\begin{document}

\section{Introduction}
The observation of the neutrinoless double beta ($0\nu \beta \beta$) decay would establish violation of lepton number conservation and the Majorana character of neutrinos. Experiments searching for the $0\nu \beta \beta$ decay are therefore considered as one of the most important experimental efforts in neutrino physics and the next generation of experiments strive to reach a sensitivity for the effective neutrino Majorana mass of $\sim 10\,\text{meV}$. The key to the success of next-generation $0\nu \beta \beta$ experiments is the efficient suppression of background events while retaining a high signal acceptance. 
In the source-equal-to-detector approach, novel experimental concepts encompass the detection of two signals, one coming from the main detector containing the $0\nu \beta \beta$ isotope and the other one from an auxiliary device, e.g., the charge signal in germanium detectors in anti-coincidence with a liquid argon scintillation veto \cite{Ref:01} or the simultaneous measurement of a heat signal and the emitted scintillation light with cryogenic detectors \cite{Ref:02,Ref:03}.
The latter technique is currently being used by the direct dark matter search experiment CRESST-II (Cryogenic Rare Event Search with Superconducting Thermometers) \cite{Ref:25,Ref:04} to distinguish different kinds of interacting particles in $\text{CaWO}_{4}$ crystals by their respective light yield using two low-temperature detectors.\\

This technique can also be applied to experiments using non-scintillating tellurium dioxide ($\text{TeO}_{2}$) crystals in the search for the $0\nu \beta \beta$ decay of $^{130}\text{Te}$ ($Q_{\beta \beta} = 2.53 \,\text{MeV}$) as proposed in \cite{Ref:06}. The $\alpha$ background in such an experiment could be discriminated from $0\nu \beta \beta $ signal events by detecting the Cherenkov light produced by the two electrons emitted in this process.
In $\text{TeO}_{2}$, the threshold for the production of Cherenkov light is $\sim 50 \text{ keV}$ for electrons and $\sim 400 \text{ MeV}$ for $\alpha$ particles. Thus, for $\alpha$ particles no light is being generated in the energy range of interest at $Q_{\beta \beta}$. 
Though the emitted $\beta$ particles are above this threshold, the amount of light being produced is very small. In the spectral range relevant for the measurement presented in this work\footnote{We consider a spectral range from $\lambda \approx 390\,\text{nm}$, where the reflectivity of the reflective foil surrounding the detector module is cut-off \cite{Ref:07}, to $\lambda \approx 1000\,\text{nm}$, where the absorption of the silicon absorber exhibits a cut-off \cite{Ref:08} (see experimental setup).}, a total energy of only $\mathcal{O}(400\,\text{eV})$ is carried by the Cherenkov photons. Previous measurements with cryogenic light detectors consisting of germanium absorbers, read out with neutron transmutation doped germanium thermistors (NTDs) \cite{Ref:26,Ref:09}, successfully showed that the detection of Cherenkov radiation from $\text{TeO}_{2}$ crystals is in principle possible. However, in \cite{Ref:26,Ref:09}, the sensitivity of the cryogenic light detectors was not sufficient to provide an effective event-by-event suppression which requires a high suppression efficiency of $\alpha$ induced events and a large signal acceptance simultaneously.\\

In this work we present the results of a measurement performed with a cryogenic light detector with transition edge sensor (TES) read-out, amplified by the Neganov-Luke (NL) effect \cite{Ref:10,Ref:11}. Such devices are currently being developed in the framework of the direct dark matter search experiments CRESST-II \cite{Ref:04} and EURECA (European Underground Rare Event Calorimeter Array) \cite{Ref:12} to investigate experimental techniques capable of further increasing the sensitivity of cryogenic light detectors at low energies. 
By employing the NL effect, the heat signal of particle interactions in a semiconductor absorber operated at cryogenic temperatures ($\mathcal{O}(\text{mK})$) can be amplified by drifting the created electrons and holes in an electric field applied to the absorber \cite{Ref:11}. For optical photons, the resulting theoretical thermal gain is given by
\begin{equation}
G = 1+\frac{e\cdot V_{NL}}{E_{ph}/\eta}
\label{eq:1}
\end{equation}
where $e$ is the electron charge, $V_{NL}$ the applied NL voltage, $E_{ph}$ the photon energy and $\eta$, the corresponding quantum efficiency for electron-hole pair production in the absorber material \cite{Ref:13}. In the spectral range relevant for the measurements presented in the present work, $\eta$ is close to unity \cite{Ref:14}.
By taking advantage of the NL effect, the threshold of a cryogenic light detector as well as the detector resolution at energies below $1-2\,\text{keV}$ can significantly be improved \cite{Ref:15}.

\section{Experimental Setup}
\begin{figure}[tbp]
\centering
\includegraphics[width=0.5\textwidth]{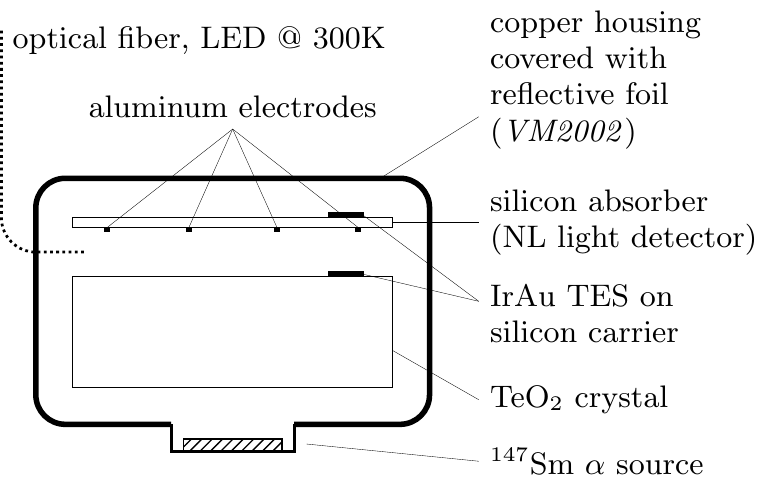}

\caption{Schematic drawing of the low-temperature detectors inside a copper housing.}
\label{fig:0}
\end{figure}

A detector module consisting of a NL light detector and a $\text{TeO}_{2}$ bolometer was operated in a dilution refrigerator in the shallow underground laboratory (overburden  $\sim 15 \,\text{ m.w.e.}$) at the Physik-Department of the Technische Universit\"at M\"unchen, Germany. A schematic drawing of the detector module is shown in fig. \ref{fig:0}.
Both detectors are mounted inside a copper housing, the inner surface of which is covered with a reflective foil (\textit{VM2002}) to increase the light collection efficiency. A $20 \times 20 \times 10 \,\text{mm}^{3}$ $\text{TeO}_{2}$ crystal with a mass of $\sim23\,\text{g}$ is mounted inside the housing using Teflon (PTFE) clamps. All sides of the crystal are roughened to reduce the amount of light trapped inside the crystal \cite{Ref:16}. Since all sides are roughened, the IrAu transition edge sensor (TES) \cite{Ref:17} is deposited onto a silicon carrier substrate ($5 \times 3 \times 0.5\, \,\text{mm}^{3}$) and then glued to the crystal. The NL light detector is a $20 \times 20 \times 0.5\,\textrm{mm}^{3}$ high purity silicon light absorber with 4 aluminum electrode strips ($18 \times 0.2 \,\text{mm}^{2}$ each, $6\,\text{mm}$ separation between strips), deposited onto the side facing the crystal.
In order to maintain electrical insulation between the TES and the silicon absorber, a silicon TES carrier is glued to the absorber. For the measurements presented in this work, a NL voltage of $V_{NL} = 70\,\text{V}$ is applied between electrodes adjacent to each other.
The $\text{TeO}_{2}$ crystal is irradiated with $\alpha$ particles from a $^{147}\text{Sm}$ ($E_{\alpha}= 2.3\,\text{MeV}$) source consisting of a sheet  of natural samarium ($10\times 10 \times 1 \text{ mm}^{3}$). Since the $\alpha$ particles are emitted throughout the source, a flat energy spectrum of degraded $\alpha$ particles is observed (expected rate $\mathcal{O}(0.1\,\text{Hz})$).\\

\section{Calibration of the Light Detector}
\begin{figure}[tbp]
\centering
\includegraphics[width=0.55\textwidth]{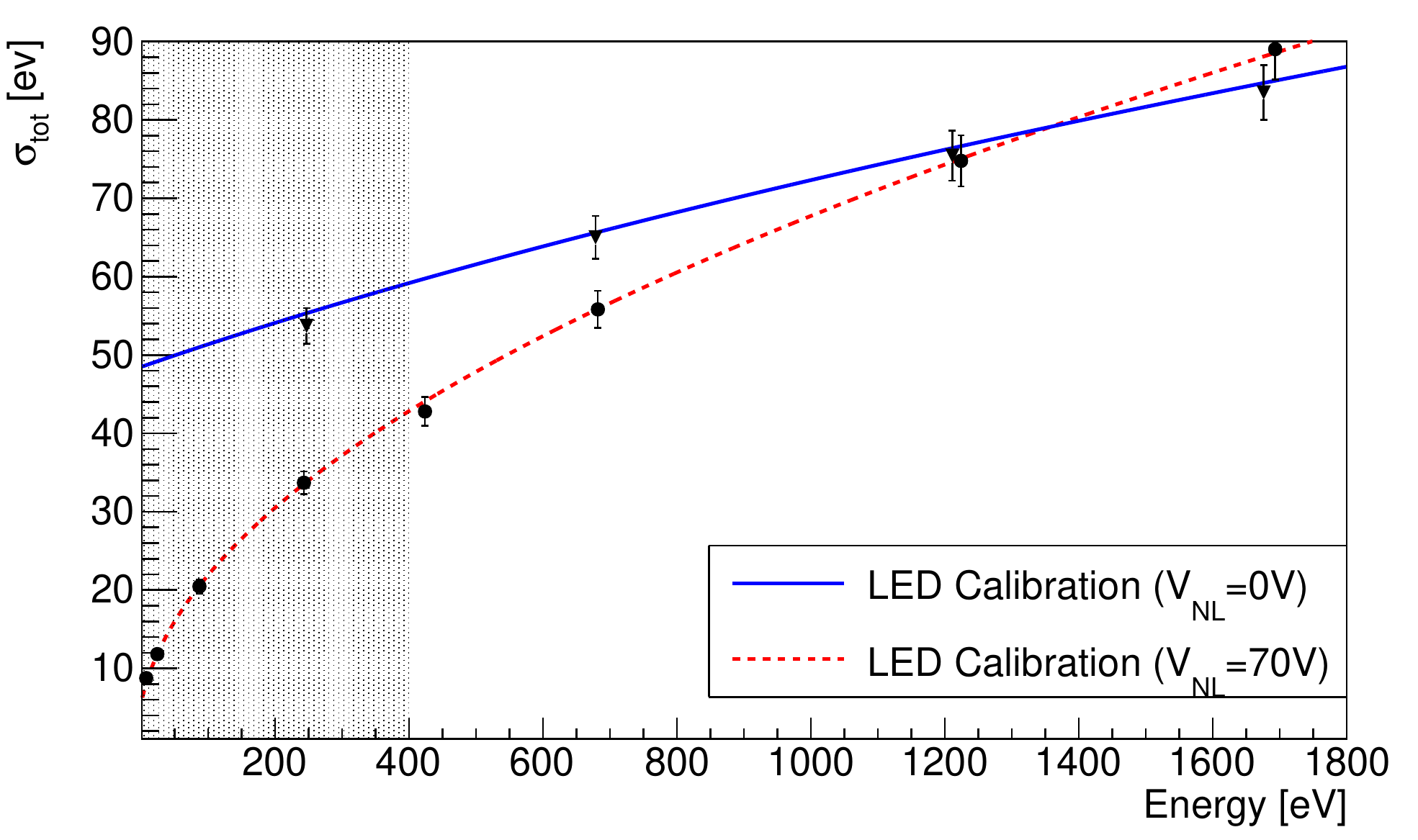}
\caption{Data points recorded during the LED calibration ($\lambda_{LED}\approx 430\,\text{nm}$) with $V_{NL}=0V$ (solid blue line) and $V_{NL}=70V$ (dotted red line). In the energy range relevant for this work (gray-shaded area, $0-400\,\text{eV}$) the resolution with applied NL voltage is significantly improved. The error bars represent the $2\sigma$ errors.}
\label{fig:1}
\end{figure}

Prior to each measurement, the light detector is calibrated using a novel calibration scheme based on photon counting statistics \cite{Ref:18}. Short light pulses ($\sim250\,\text{ns}$) of varying intensity are generated by a light-emitting diode (LED) operated at room temperature and are guided onto the NL detector using an optical fiber (see fig. \ref{fig:0}). Since the development of these detectors is performed within the framework of direct dark matter search with $\text{CaWO}_{4}$ crystals, a LED matching the scintillation spectrum of these crystals is used ($\lambda_{LED} \approx 430\,\text{nm}$).
With applied NL voltage, the observed gain is  $G_{ob}=10.8\pm0.1$ and an improvement in the signal-to-noise (S/N) ratio of $6.2 \pm 0.1$ is achieved. 
The difference between the gain and the improvement of the S/N ratio is due to additional electronic noise introduced by applying the NL-voltage.
The theoretical gain predicted by eq. (\ref{eq:1}) is $\sim 25$ and the discrepancy to the observed value is due to a reduced drift length of the charge carriers, caused by trapping in impurities and defects at the absorber surface \cite{Ref:18}. When the light detector is operated without an applied NL voltage ($V_{NL}=0\,\text{V}$), the energy resolution is described by the following equation \cite{Ref:18}
\begin{equation}
\sigma_{tot}=\sqrt{\sigma_{0}^{2}+\sigma_{ph}^{2}}=\sqrt{\sigma_{0}^{2}+E_{ph} \cdot E}
\label{eq:2}
\end{equation}
where $\sigma_{0}$ refers to all energy-independent contributions to the detector resolution, $\sigma_{ph}$ describes the contribution due to photon counting statistics, and $E$ is the energy detected by the light detector. Contributions of higher order are neglected. Here, the energy-independent term was determined to be $\sigma_{0}=48.5\pm1.1\,\text{eV}$.
When operated with an applied NL voltage, an extended function which takes higher-order terms into account is used to describe the detector resolution \cite{Ref:15,Ref:21}:
\begin{equation}
\sigma_{tot} = \sqrt{\sigma_{NL}^{2}+E_{ph} \cdot E + a_{cc}\cdot E + b_{rc}\cdot E^{2}}
\label{eq:3}
\end{equation}
The term $\sigma_{NL}$ describes the energy-independent contributions to the resolution when operated with applied NL voltage. The parameters $a_{cc}$ and $b_{rc}$ are physically motivated parameters \cite{Ref:15} which take into account incomplete charge collection and charge recombination processes, respectively. These additional parameters increase the uncertainty of the detected energy and are device-specific and depend on the applied voltage. With applied NL voltage, the energy-independent term was found to be $\sigma_{NL}=7.8\pm 0.2\,\text{eV}$. The results of both calibration measurements are shown in fig. \ref{fig:1}. In the energy range relevant for the detection of Cherenkov light emitted by $\text{TeO}_{2}$ (gray-shaded area  in fig. \ref{fig:1}, $0-400\,\text{eV}$), the resolution with applied NL voltage is significantly improved while above $\sim 1.4 \,\text{keV}$, the resolution is deteriorated.
When investigating Cherenkov light emitted from a $\text{TeO}_{2}$ crystal, the actual light detector resolution can differ from the resolution determined by the LED calibration
since the mean wavelength and the spectral shape of the detected photons are expected to be different from the wavelength and spectral shape of the LED used.
For this special case, the detector resolution still has to be studied in further detail.
However, an exact knowledge of the energy dependence of the light detector resolution as well as an absolute energy calibration is not required to determine the suppression efficiency of $\alpha$ from $\text{e}^{-}/\gamma$ events in the measured spectra.

\section{Results}
\begin{figure}[tbp]
\centering
\includegraphics[width=0.6\textwidth]{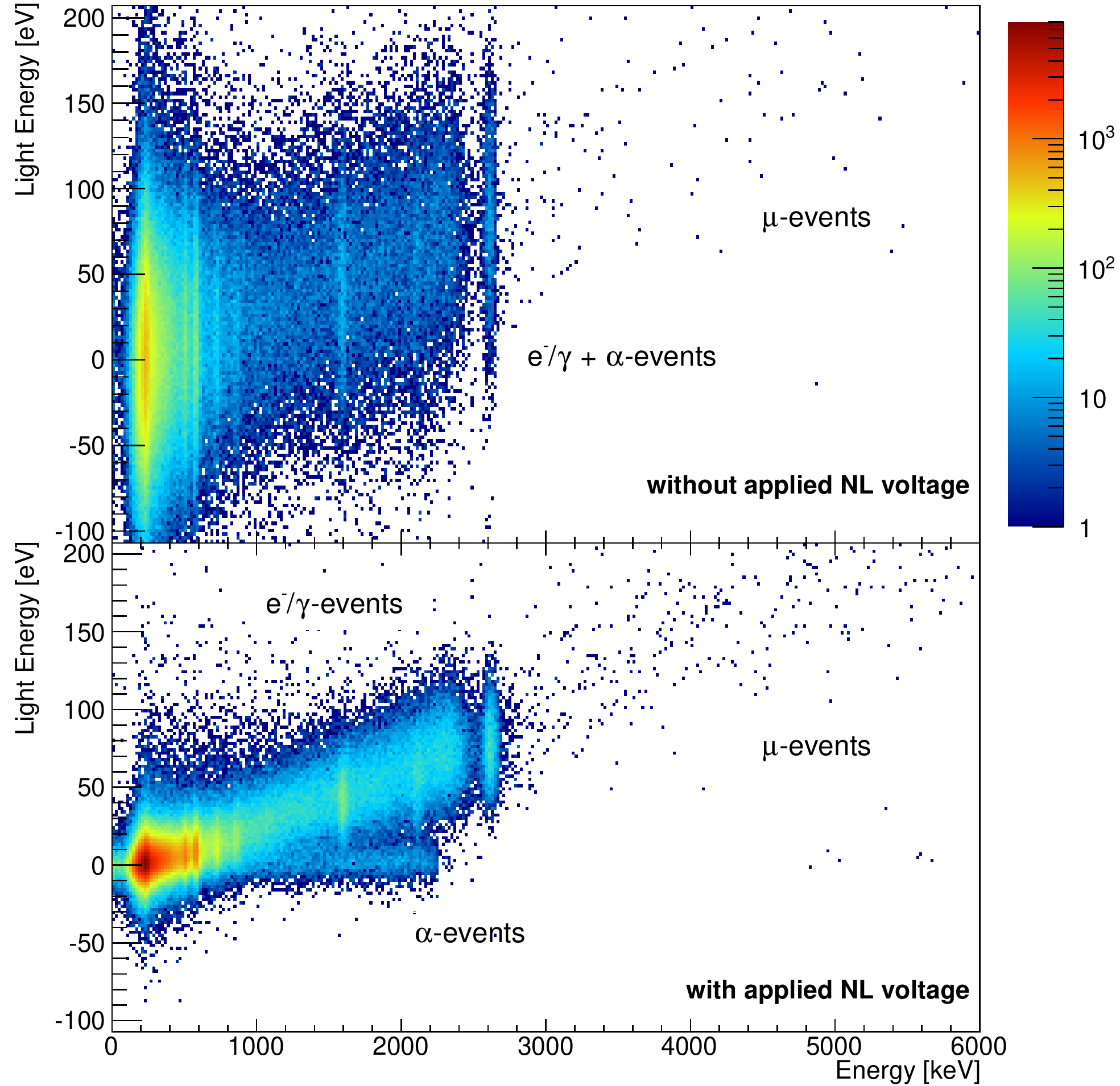}
\caption{
Two-dimensional histogram of the measurements performed without (top) and with applied NL voltage (bottom). In the measurements, events induced by the $^{228}\text{Th}$ source (inclined band), by the $^{147}\text{Sm}$ $\alpha$-source (horizontal band, only visible with applied NL voltage) and muon-induced events (inclined band above $\sim 2.6\,\text{MeV}$) are visible.
}
\label{fig:2}
\end{figure}

Two measurements with a $^{228}\text{Th}$ $\gamma$-source (rate $\sim 5 \,\text{Hz}$), placed outside the cryostat, were performed in order to determine the discrimination efficiency between $\alpha$ and $\text{e}^{-}/\gamma$-induced events. 
A measurement ($\sim 16 \,\text{h}$) without applied NL voltage (fig. \ref{fig:2} top) and a measurement ($\sim 42\,\text{h}$) with applied NL voltage (fig. \ref{fig:2} bottom) were performed.
The energy calibration of the light detector is based on the results of the LED calibration \cite{Ref:15,Ref:18,Ref:21}, the calibration of the $\text{TeO}_{2}$ detector is performed using $\gamma$-peaks from the $^{228}\text{Th}$ source.
In both spectra, $\text{e}^{-}/\gamma$ events (inclined band) and $\alpha$ events (horizontal band) are present. However, the population of $\alpha$-induced events can only be distinguished from $\text{e}^{-}/\gamma$ events when the NL voltage is applied to the detector. Both measurements also include events induced by atmospheric muons which can be identified at energies $> 2.6\,\text{MeV}$.
At the full energy peak of $^{208}\text{Tl}$ ($2.614\,\text{MeV}$), an energy resolution of $\Delta E_{FWHM} = 67\,\text{keV}$ is achieved in the phonon detector. The observed resolution is worse by a factor of $\sim 10$ than the resolution achievable with CRESST-type cryogenic detectors with TES read-out ($\Delta E_{FWHM} = 6.7 \,\text{keV}$ at $2.31\,\text{MeV}$) \cite{Ref:19}. This is probably caused by an acoustic mismatch between the silicon TES carrier and the $\text{TeO}_{2}$ crystal or due to a poor glue-interface between the crystal and the carrier. Since only one $\text{TeO}_{2}$ crystal was available for the measurements performed in the present work, no further tests concerning the energy resolution could be performed.\\ 

As reported in \cite{Ref:18}, NL light detectors, when operated with an applied NL voltage, can exhibit a reduction of the observed gain with time due to an accumulation of charge carriers in the vicinity of the aluminum electrodes. This causes an electric field with reverse polarity with respect to the applied voltage and therefore reduces the effective NL voltage. In the measurement presented here, a reduction of the gain by $\sim 10\%$ during a $42\,\text{h}$ measurement is observed.

A simulation performed with Geant4 \cite{Ref:20} shows that the total amount of Cherenkov light produced for the two electrons emitted in the $0\nu \beta \beta$ decay is practically identical to the amount of light produced for $\gamma$ events in the full energy peak (FEP) of $^{208}\text{Tl}$ \cite{Ref:21}. Therefore, we determine the signal in the light detector expected at $Q_{\beta \beta}$($^{130}\text{Te}$) using events from the FEP of $^{208}\text{Tl}$. 
The signal in the light detector for $\alpha$-induced events is obtained in a reference region between 2.0 and 2.3 MeV and is assumed to be also valid at $Q_{\beta \beta}$. 
This has also been supported by independent measurements which show that no Cherenkov light is being produced for $\alpha$-induced events in TeO$_{2}$ \cite{Ref:09}.

\begin{figure}[tbp]
\centering
\includegraphics[width=0.6\textwidth]{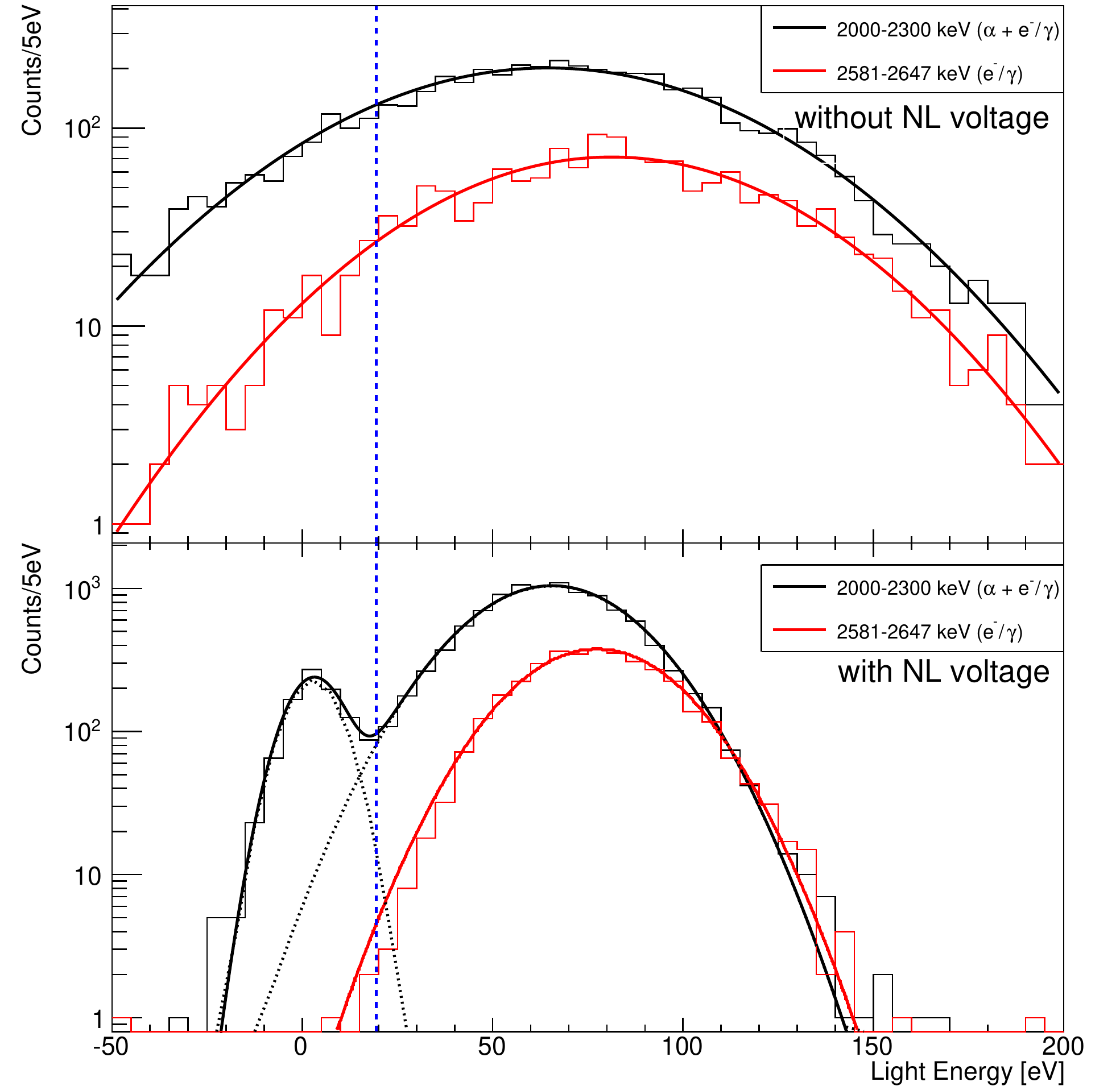}
\caption{Logarithmic histogram of the light energy for $\gamma$ events from the full energy peak (FEP) of $^{208}\text{Tl}$ (red) and $\alpha$ \& $\gamma$ events obtained from a reference region between 2.0 and 2.3 MeV (black) for the measurement without applied NL voltage (top) and with applied NL voltage (bottom). The plots also show the normal distributions fitted to the individual histograms. The vertical (dashed, blue) line indicates the signal acceptance of $99.8\,\%$ in the full energy peak when operated with applied NL voltage.}
\label{fig:3}
\end{figure}

A histogram showing the individual light-energy distributions is depicted in fig. \ref{fig:3}. We assume that all contributions can be described by normal distributions. For the measurement without applied NL voltage, the light-energy distribution of both reference regions is fitted using a single normal distribution (fig. \ref{fig:3} top), while for the measurement with applied NL voltage, the light-energy distribution of the $\alpha$ reference region is fitted with two normal distributions (fig. \ref{fig:3} bottom), accounting for $\alpha$ as well as $\text{e}^{-}/\gamma$ events. The individual fits are characterized by the mean $x_{0}$ and the variance $\sigma$ of the distribution. The results of the individual fits are given in table \ref{tab:1} and directly show the improved energy resolution of the light detector when operated with applied NL voltage. Due to a small reduction of the gain throughout the measurement (mentioned above), the mean light energy in the FEP of $\text{e}^{-}/\gamma$ events is slightly reduced in the measurement performed with NL voltage compared to the measurement without applied voltage.\\

In the measurement without applied NL voltage, the different contributions to the $\alpha$ reference region cannot be disentangled and no discrimination of $\alpha$-induced events is possible. However, in the measurement performed with applied NL voltage, the individual distributions in the $\alpha$ reference region can be clearly identified (dotted black lines in bottom panel of fig. \ref{fig:3}). The contribution of the $\alpha$ events can be clearly separated from the $\text{e}^{-}/\gamma$ events in the full energy peak\footnote{With applied NL voltage, the mean light energy $x_{0}$ of the $\alpha$ events is $\sim 2.9\,\text{eV}$ (see tab. \ref{tab:1}). This systematic, energy-independent offset is present when the light detector is operated with applied NL voltage and is caused by a small cross-talk between both detectors. The width of the distribution of $\alpha$-induced events is compatible with the noise present in the light detector, indicating that, in fact, no light is being detected.}.
We determine the efficiency of the $\alpha$-suppression by requiring a high signal acceptance in the full energy peak, i.e. a threshold in the light detector is chosen in such a way that $99.8\%$ of all events in the FEP are accepted, which leads to a threshold of 19.0 eV (blue line in fig. \ref{fig:3}). From the distribution of the $\alpha$ events in the $\alpha$ reference region it follows that the amount of $\alpha$ events above this threshold is $\sim1\%$. These results show that, for energies close to $Q_{\beta \beta}(^{130}\text{Te})$, $\text{e}^{-}/\gamma$ events can be discriminated from $\alpha$-induced background events on an event-by-event basis.\\

\begin{table}[tbp]
\caption{Values obtained for the mean $x_{0}$ and the standard deviation $\sigma$ of individual normal fits shown in figure~4.}
\label{tab:1}
\smallskip
\centering
\begin{tabular}{ l c c c }
\hline\noalign{\smallskip}
& $E$ [keV] & $x_{0}$ [eV] & $\sigma$ [eV]\\
\noalign{\smallskip}\hline\noalign{\smallskip}
\multicolumn{4}{c}{$V_{NL}=0V$}\\

$\alpha + \text{e}^{-}/\gamma$ & $2000 - 2300$ & $64.6\pm0.7$ & $48.8\pm0.6$ \\
$\text{e}^{-}/\gamma$\,(FEP)   & $2581 - 2647$& $81.2\pm0.9$ & $44.0\pm0.9$ \\

\noalign{\smallskip}\hline\noalign{\smallskip}

\multicolumn{4}{c}{$V_{NL}=70V$}\\

$\alpha$ & $2000 - 2300$ & $2.9 \pm 0.3$ & $7.1\pm0.3$ \\
$\text{e}^{-}/\gamma$ & $2000 - 2300$ & $65.6\pm0.2$ & $20.4\pm0.2$ \\
$\text{e}^{-}/\gamma$\,(FEP)   & $2581 - 2647$& $77.5\pm0.3$ & $19.5\pm0.2$ \\

\noalign{\smallskip}\hline
\end{tabular}

\end{table}
The sensitivities predicted for next-generation experiments searching for the neutrinoless double beta decay of $^{130}\text{Te}$ with $\text{TeO}_{2}$ based cryogenic bolometers crucially depend on the background level in the region of interest (ROI) around $Q_{\beta \beta}$ \cite{Ref:22}.
A significant reduction of the background level is key to overcoming this limiting factor of the sensitivity.
The CUORE-0 (Cryogenic Underground Observatory for Rare Events) experiment \cite{Ref:23}, employing $\text{TeO}_{2}$ based cryogenic detectors, reports background levels of $0.019 \,\text{cts}/\text{keV}/\text{kg}/\text{yr}$ in the flat $\alpha$ continuum region between 2.7 MeV and 3.4 MeV. This value allows to anticipate a background level for CUORE \cite{Ref:24} of the order of $0.01 \,\text{cts}/\text{keV}/\text{kg}/\text{yr}$. The results presented here show that a further $\alpha$ background reduction of about two orders of magnitude is achievable using the phonon-light technique with NL amplified cryogenic light detectors. Therefore, the contribution of the $\alpha$ background in the ROI around $Q_{\beta \beta}$ in next-generation $\text{TeO}_{2}$ experiments implementing this technology is compatible with a background index of $10^{-4}$ cts/keV/kg/yr, which corresponds to a total $\alpha$ background close to zero at the ton$\times$year exposure scale, given the excellent energy resolution already achieved by $\text{TeO}_{2}$ bolometers \cite{Ref:23}.\\

The background suppression technique described in the present work can provide an important contribution to future experiments by providing the ability to actively discriminate $\alpha$ induced background from $\text{e}^{-}/\gamma$ events with a high efficiency, while at the same time, retaining a high signal acceptance. 
However, an important aspect, which has to be considered by the individual experiments when incorporating such light detectors in future-generation searches, is a possible introduction of additional backgrounds into the experiment, especially due to the additional wiring required to operate the detectors.
Another aspect to be taken into account is related to the crystal size. Due to practical reasons, the crystals used in bolometric arrays searching for $0 \nu \beta \beta$ have masses in the $500\,\text{g} - 1000\,\text{g}$ range and it is therefore important to show in future tests that the suppression factor achieved in this work can also be reached with such larger crystals. Recent measurements have shown that a background suppression as demonstrated in the present work can also be achieved in larger $\text{TeO}_{2}$ crystals (mass $\sim280\,\text{g}$) \cite{Ref:30}. For future systematic studies of the background suppression efficiency, the light collection efficiency in the detector module and the dependence on the optical properties of TeO$_{2}$ crystals (e.g. the absorption and scattering lengths) and crystal shape has to be studied in further detail. Simulations performed in \cite{Ref:21} show that the detectable amount of Cherenkov radiation produced in a TeO$_{2}$ crystal could be increased by a factor of $\sim 2$ by using a reflector foil with a higher reflectivity in the UV compared to the presently used VM2002 foil. Furthermore, recent work \cite{Ref:15,Ref:21} indicates that light detectors based on NL technology can be improved even further, in particular with respect to a reduction of the thermal gain with time and the achievable thermal gain and, therefore, to S/N ratio.

\section{Conclusion}
We demonstrate for the first time, that an event-by-event discrimination between $\text{e}^{-}/\gamma$ and $\alpha$-induced events at $Q_{\beta \beta}(^{130}\text{Te})$ in $\text{TeO}_{2}$ based bolometers is possible using NL  light detectors. We reach an $\alpha$-suppression of $99\%$ while accepting $99.8\%$ of $\text{e}^{-}/\gamma$ events at the full energy peak of $^{208}\text{Tl}$. The achieved suppression factor could be further improved by an increased light collection efficiency of the detector module (due to the presence of the Sm source and to accommodate electrical and mechanical feed-throughs, the detector housing was not completely covered with the reflective foil \cite{Ref:21}) and by further enhancing the performance of the NL light detector, in particular concerning the achieved gain and the electronic noise introduced by the application of the NL voltage.

\acknowledgments
We gratefully acknowledge the support of the DFG cluster of excellence ``Origin and Structure of the Universe'', the ``Helmholtz Alliance for Astroparticle Phyiscs'', and the ``Maier-Leibnitz-Laboratorium'', Garching.

\end{document}